\documentclass[sigconf]{acmart}
\AtBeginDocument{%
  \providecommand\BibTeX{{%
    \normalfont B\kern-0.5em{\scshape i\kern-0.25em b}\kern-0.8em\TeX}}}

\setcopyright{acmcopyright}
\copyrightyear{2024}
\acmYear{2024}
\setcopyright{rightsretained}
\acmConference[LAK '24]{The 14th Learning Analytics and Knowledge Conference}{March 18--22, 2024}{Kyoto, Japan}
\acmBooktitle{The 14th Learning Analytics and Knowledge Conference (LAK '24), March 18--22, 2024, Kyoto, Japan}\acmDOI{10.1145/3636555.3636883}
\acmISBN{979-8-4007-1618-8/24/03}

\usepackage{acmart-taps}

\usepackage{svg}
\usepackage{mathtools}
\usepackage{bm}
\usepackage{xspace}
\usepackage{multirow}
\usepackage{soul}

\newlength{\dss}
\newcommand{\calE}{\mathcal{E}}

\newcommand{\calL}{\mathcal{L}}

\newcommand{\calT}{\mathcal{T}}

\newcommand{\bbR}{\mathbb{R}}

\DeclareMathOperator*{\argmax}{arg\,max}

\DeclarePairedDelimiter{\norm}{\lVert}{\rVert}

\DeclarePairedDelimiterX{\ip}[2]{\langle}{\rangle}{#1,#2}

\newcommand{\methodname}{\text{mHyER}\xspace}
\newcommand{\ex}[1]{e_{#1}}
\newcommand{\exh}[1]{\tilde{e}_{#1}}
\newcommand{\exfl}[1]{\ex{#1}^{(\texttt{L1})}}
\newcommand{\exll}[1]{\ex{#1}^{(\texttt{L2})}}

\newcommand{\encoder}{f_{\theta}}

\newcommand{\duord}{\text{DuoRD dataset}\xspace}
\newcommand{\duolingo}{\text{Duolingo}\xspace}

\newcommand{\bert}{\text{BERT}\xspace}
\newcommand{\mbert}{\text{mBERT}\xspace}
\newcommand{\contriever}{\text{Contriever}\xspace}
\newcommand{\mcontriever}{\text{mContriever}\xspace}
\newcommand{\simcse}{\text{SimCSE}\xspace}
\newtoggle{comments}
\togglefalse{comments}

\iftoggle{comments}{%
    
    \newcommand{\red}[1]{\textcolor{red}{#1}}

    \newcommand{\axcomment}[1]{{\bf{{\red{{#1}}}}}}
}{%
    \newcommand{\axcomment}[1]{}

}

\usepackage{lipsum}

\begin{document}

\title{Large Language Model Augmented Exercise Retrieval for Personalized Language Learning}

\author{Austin Xu}
\authornote{Work done during an internship at Duolingo. Contact: \texttt{axu@gatech.edu}}
\affiliation{%
  \institution{Georgia Institute of Technology}
  \city{Atlanta}
  \state{Georgia}
  \country{USA}
}
\email{axu@gatech.edu}

\author{Will Monroe}
\affiliation{%
  \institution{Duolingo}
  \city{Pittsburgh}
  \country{USA}  
}
\email{monroe@duolingo.com}

\author{Klinton Bicknell}
\affiliation{%
  \institution{Duolingo}
  \city{Pittsburgh}
  \country{USA}  
}
\email{klinton@duolingo.com}

\renewcommand{\shortauthors}{Xu, Monroe, and Bicknell}

\begin{abstract}
  We study the problem of zero-shot exercise retrieval in the context of online language learning, to give learners the ability to explicitly request personalized exercises via natural language. Using real-world data collected from language learners, we observe that vector similarity approaches poorly capture the relationship between exercise content and the language that learners use to express what they want to learn. This semantic gap between queries and content dramatically reduces the effectiveness of general-purpose retrieval models pretrained on large scale information retrieval datasets like MS MARCO~\cite{bajaj2016ms}. We leverage the generative capabilities of large language models to bridge the gap by synthesizing hypothetical exercises based on the learner's input, which are then used to search for relevant exercises. Our approach, which we call \methodname, overcomes three challenges: (1) lack of relevance labels for training, (2) unrestricted learner input content, and (3) low semantic similarity between input and retrieval candidates. \methodname outperforms several strong baselines on two novel benchmarks created from crowdsourced data and publicly available data.
\end{abstract}




\maketitle

\section{Introduction}\label{sect:intro}
Modern personalized education systems typically leverage the power of machine learning models to estimate user skill levels~\cite{corbett1994knowledge} and adaptively serve exercises to learners~\cite{wu2020exercise, huang2022design, cui2023adaptive}.
Adaptivity, while a critical part of any personalized education system, is a \emph{passive} form of personalization from the learner's point of view: While exercises are tailored to an estimate of the learner's skill level, this customization occurs behind the scenes, with no opportunity for learners to take initiative in shaping the learning process. In this paper, we study a complementary form of \emph{learner initiated} personalization in the context of \emph{online language learning}. In particular, learners are given the ability to \emph{explicitly} request learning content from an education system, which returns relevant exercises from a fixed catalog for the learner to do. 

\begin{figure}[t!]
    \centering
    \includegraphics[width=0.8\linewidth]{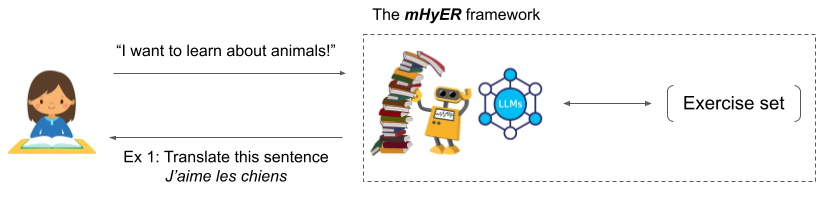}
    \caption{Exercise retrieval for learner directed language learning and our proposed solution, multilingual Hypothetical Exercise Retriever (\methodname). At a high level, learners are allowed to provide \emph{any} natural language input, and the goal is to retrieve exercises relevant to that input. Our method utilizes large language models to perform zero-shot retrieval.}
    \label{fig:teaser}
\end{figure}

This type of learner initiated personalization can be viewed as a form of \emph{self-directed learning}, where learners take initiative over the learning process. Self-directed learning has been shown to increase learner performance across multiple topics~\cite{mandasari2021flipped, lucieer2016self, kim2015changes, krashen2005free}, improve learner motivation~\cite{patall2008effects}, and create more cohesive learner environments~\cite{geng2019investigating}. Online language learning is a natural setting for self-directed learning, as people learn languages for very personal reasons: Some learn for fun, while others have specific goals, such as preparing for an international trip or developing language skills for business. Different reasons for learning lead to different needs for exercise content: Someone learning to write in a business setting may want extra practice with grammar or politeness, whereas the learner learning for a vacation may want exercises about hotels or transportation. Beyond highly personalized motivations for learning, online language learners do not have immediate access to instructors who can plan learning material to target weaknesses. As such, there is an inherent need for online language learners to have some degree of self-direction in order to get the most out of their learning experience.

With the goal of allowing language learners to tailor an online learning experience to their own needs, we formalize the task of \textbf{exercise retrieval for learner directed language learning} and evaluate machine learning models for this task. The goal of this task (Figure~\ref{fig:teaser}) is to retrieve relevant exercises from a set of existing exercises based on a learner's input. In this setting, collecting relevance labels (i.e., pairs of learner inputs and relevant exercises) is particularly challenging, as learners will typically be presented with only a small number of exercises for any given input. As a result, we consider the \emph{zero-shot} setting, where we do not have access to relevance labels for training. While many off-the-shelf models exist for text-based retrieval, we show that direct similarity search (i.e., retrieving exercises that are the most similar to the user input in the representation space) with these models suffers from a semantic similarity gap between how users describe their learning objectives and exercise content. To overcome this gap, we leverage structure inherent to exercises and the generative capabilities of large language models. Specifically, we make the following contributions.
\begin{itemize}
    \item We propose the new task of exercise retrieval for learner directed language learning in Section~\ref{sec:prob_setup} and discuss how learner inputs give rise to a fundamental challenge in this task.
    \item We present our zero-shot retrieval approach, \methodname, in Section~\ref{sec:method}, and illustrate how augmenting retrieval with LLMs helps overcome the pitfalls of direct similarity search.
    \item With no existing benchmarks for this task, we create two novel benchmarks in exercise retrieval with both crowd-sourced data from learners of a popular language learning app and publicly available Tatoeba data. We evaluate our method against several strong dense retrieval baselines in Section~\ref{sec:exp} and empirically show that \methodname outperforms relevant baselines by a significant margin.
\end{itemize}

\subsection{Related work}
Exercise retrieval is naturally connected to the broad field of information retrieval, and in particular, dense retrieval~\cite{lee2019latent,karpukhin2020dense}. Dense retrieval focuses on retrieving documents based on similarity measured in a learned representation space. Zero-shot retrieval, or retrieval without training on task-specific relevance information, is of particular relevance to our task. Such methods typically rely on a supervised pretraining stage~\cite{thakur2021beir,wang2022gpl,yu2022coco}, where models are trained on large scale retrieval datasets, such as MS MARCO~\cite{bajaj2016ms}. However, such supervised pretraining ultimately depends on the existence of suitable labeled datasets, which are not always readily available~\cite{izacard2022unsupervised}. The rise of large language models (LLMs) with strong zero/few-shot performance in new domains has resulted in a line of research integrating LLMs into the retrieval pipeline. Such approaches typically rely on some combination of specialized prompting and synthesizing retrieval datasets to retrain retrieval models~\cite{bonifacio2022inpars, dai2022promptagator, sachan2022improving, yu2022generate}. Our approach takes particular inspiration from HyDE~\cite{gao2023precise}, which utilizes a LLM to synthesize a hypothetical document, which is used then used with a pretrained encoder to retrieve documents via nearest neighbors. 

A fundamental step in any retrieval method is the representation space used for similarity comparisons. For the task of exercises retrieval, we focus on learning sentence embeddings, where pretrained language models such as \bert~\cite{devlin2019bert} or RoBERTa~\cite{liu2019roberta} serve as strong foundations. Contrastively learning sentence representations, which leverage techniques used in the image domain~\cite{hadsell2006dimensionality,chen2020simple}, has become especially popular. The goal of contrastive learning is to learn a representation space where similar items (``positive pairs'') are pulled close together while dissimilar items (``negative pairs'') are pulled far apart in an unsupervised manner. In the image domain, positive pairs are formed by applying \emph{data augmentation}, such as cropping or rotating an image. Such techniques are not directly transferable to natural language, resulting in a long line of methods~\cite{izacard2022unsupervised, gao2021simcse, chuang2022diffcse, wu2022esimcse, cheng2023improving} studying contrastive sentence embeddings. Of particular interest to the language learning setting is multilingual contrastive learning~\cite{wang2022english}, where positive pairs can be taken as the same sentence in two different languages. In all, \methodname can be viewed as a combination of multilingual contrastive learning~\cite{wang2022english} and HyDE~\cite{gao2023precise}.

Personalized education systems often gauge a learner's skill level via Knowledge Tracing~\cite{corbett1994knowledge} in order to tailor exercise difficulty level. As a result, a variety of contemporary machine learning methods~\cite{piech2015deep, pandey2019self, shin2021saint+, abdelrahman2019knowledge, tong2020structure, xu2020dynamic} have been developed to track learner skill level from historical data. Such methods demonstrate strong empirical success and thus have been leveraged to adaptively recommend exercises to learners~\cite{wu2020exercise, huang2022design} or even generate new exercises based on skill level~\cite{cui2023adaptive}. This adaptivity can be viewed as a complementary piece to the problem of exercise retrieval directed language learning that we study in this paper: learner initiated personalization can leverage existing tools from adaptivity to ensure exercises are both relevant and at the right skill level. On the other hand, adaptive systems can benefit from explicit learner direction. For example, we can view the learner input of ``past tense verbs'' as the learner explicitly saying they are not comfortable with past tense verbs, and use this information in skill estimation.

\section{Problem setup}\label{sec:prob_setup}
The goal of exercise retrieval for learner directed language learning is to retrieve relevant exercises for a learner given a text input from the learner describing what they want to learn. In particular, we assume that learner is taking a language learning course, which consists of two languages: the ``first language'' (i.e., a language they already know) and the ``second language'' (i.e., the language they are learning), which we refer to as L1 and L2, respectively.\footnote{These labels are an imprecise shorthand; L1 need not be the learner's first or native language, and L2 may be a third language or beyond.} The learner completes \emph{exercises}, which are drawn from a fixed set of $N$ exercises $\calE = \{\ex{1}, \ldots, \ex{N}\}$ that are at an appropriate level for the learner. 
We can view this set of $N$ exercises as samples from an unknown \emph{exercise distribution}, which captures characteristics (style, length, content, etc) of exercises. For simplicity, we limit our attention to translation exercises, in which a learner translates an L1 sentence $\exfl{i}$ to the L2, with one correct L2 answer $\exll{i}$ available as an example of a correct translation. The learner provides some input $t$, and our objective is to retrieve the $K$ (unique) exercises that are the most relevant based on input $t$ in a zero-shot manner. That is, without using any labeled relevance data for training, we want to retrieve $K$ unique exercises $\ex{1}^\star, \ldots, \ex{K}^\star$ that maximize probability $p$ that an exercise is relevant conditioned on learner input $t$:
\begin{align}\label{eq:objective}
    \ex{1}^\star, \ldots, \ex{K}^\star = \argmax_{\substack{\ex{1}, \ldots \ex{K} \in \calE\\\ex{i} \neq \ex{j} \quad \forall i, j,\ i \neq j}}\quad \prod\limits_{i = 1}^K p(\ex{i} | T = t).
\end{align}

\subsection{Learner inputs.} 
The core of the personalized experience in this problem setting is allowing learners to provide an input describing what they want to learn with \emph{no restrictions on input content}, resulting in large number of potential input types. For example:
\begin{itemize}
    \item \textbf{Topics:} Learners can request exercises that teach vocabulary relevant to a particular topic. Inputs such as ``words about animals'' or ``countries'' are such examples.
    \item \textbf{Grammar:} Learners can request exercises teaching grammatical concepts, such as ``non-present tenses'' or ``irregular plurals''.
    \item \textbf{Culture:} Learners can request to review culture-specific aspects of language, such as idioms, slang, or region-specific quirks. For example, a learner learning Spanish may want to learn about ``voseo'', a region-specific grammatical concept in South America.
    \item \textbf{Learning process:} Learners can request exercises that help with particular parts of the process of learning a language, such as ``words that are hard to spell'' or ``sentences for first-year students''.   
\end{itemize}
Learner inputs of these types result in what we call a \emph{referential similarity gap}: Under modern text-based retrieval models, how learners express their learning objectives (i.e., the learner input $t$) is not considered similar to what it is referring to, i.e., the content of the exercises $\exfl{}$ and $\exll{}$. We explore this gap in greater detail in Section~\ref{sec:prob_setup:challenges}.  

\section{Method}\label{sec:method}
\begin{figure}[t]
    \centering
    \includegraphics[width=0.85\linewidth]{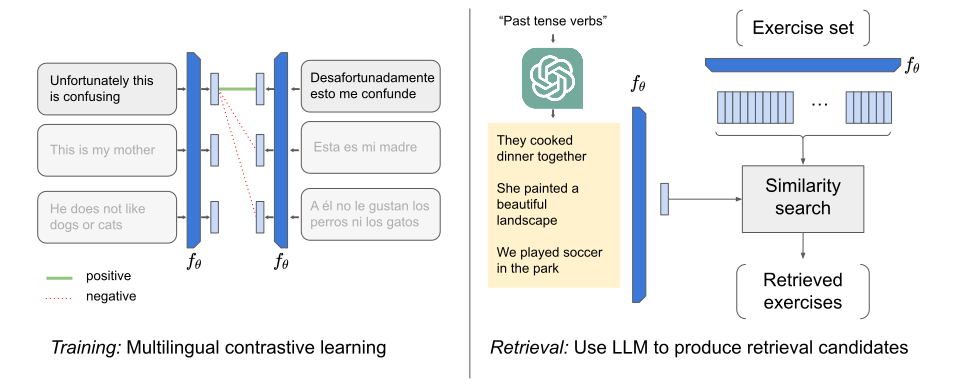}
    \caption{\methodname consists of two stages. Contrastive finetuning (left) is employed as a training stage to optimize our semantic similarity space for multilingual exercises. Then at retrieval time (right), a large language model is employed to synthesize hypothetical retrieval candidates. These retrieval candidates are then used in direct similarity search to retrieve exercises.}
    \label{fig:method}
\end{figure}

In this section, we present \underline{m}ultilingual \underline{Hy}pothetical \underline{E}xercise \underline{R}etriever (\methodname), our zero-shot exercise retrieval framework, and show that it overcomes the pitfall of direct similarity search in learned representation spaces.

\subsection{Baseline: direct search with similarity spaces.}
The backbone of text-based retrieval is a vector space representation of text that reflects some notion of similarity between different pieces of text. Forming these representation spaces remains a core part of text-based retrieval, with early methods such as BM25~\cite{robertson2009probabilistic} formed representations via word frequency. Such methods struggle to generalize as their representation spaces are formed based on counting exact or near text matches. To improve generalization, contemporary methods for text-based retrieval typically train a model $\encoder$ (parametrized by $\theta$) that maps natural language inputs (from the space of all text inputs $\calT$) to some $d$-dimensional vector space: $\encoder:\calT \rightarrow \bbR^d$. Such models are referred to as \emph{encoders}, and learn representations of text called \emph{embeddings}. That is, if $t \in \calT$ is some text, then $\encoder(t)$ is its embedding representation. Because exercises are typically short sentences or sentence fragments, we focus on encoders specifically geared towards learning sentence embeddings in this work. 

Harnessing the vast availability of text data, contemporary encoders are typically neural networks trained such that texts with similar content are more similar in the representation space under some measure, like cosine similarity. That is, if $t_1, t_2 \in \calT$ are similar in content, then $\text{sim}(\encoder(t_1), \encoder(t_2))$ is large (and positive). This similarity space suggests a natural approach for retrieving exercises: Pass each exercise $\ex{i}$ through the model $\encoder$ to obtain its embedding representation $\encoder(\ex{i})$.\footnote{
    We slightly abuse notation here and write $\encoder(\ex{i})$ to mean either $\encoder(\exfl{i})$ or $\encoder(\exll{i})$. The choice to compare against the representation of the L1 or L2 sentence is explored in Section~\ref{sec:exp}.
} Then, when a learner provides an input $t$, pass $t$ through the model to obtain $\encoder(t)$ and return the $K$ exercises with largest cosine similarity $\text{sim}(\encoder(\ex{i}),\encoder(t))$. As we see in Section~\ref{sec:prob_setup:challenges}, direct similarity search often retrieves sentences featuring ``language about language'', which are often irrelevant to the learner's input. This leads us to leverage the generative abilities of LLMs, as we discuss next.

\begin{figure*}
    \centering
    \includegraphics[width=0.9\linewidth]{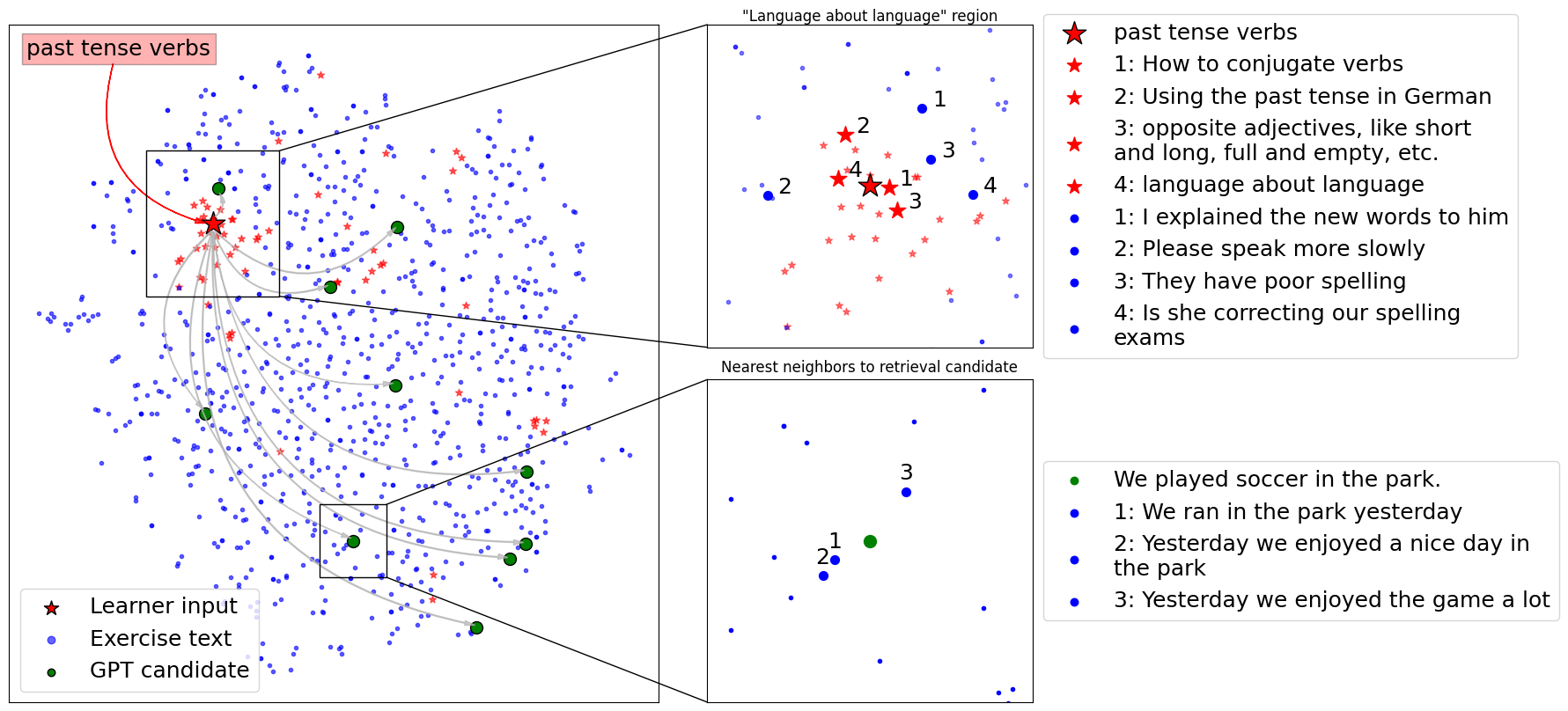}
    \caption{TSNE visualization of exercise, learner input, and GPT-4-synthesized retrieval candidate representations in the representation space of a trained \mbert encoder (left). Learner inputs concentrate in the language about language region (top right), making direct similarity search sub-optimal. Retrieval candidates bridge the referential similarity gap between learner inputs and exercise text and are close in similarity to exercises that meet the learner's specifications (bottom right).}
    \label{fig:emb-vis-final}
\end{figure*}

\subsection{\methodname: augmenting direct search with generative capabilities.}
If large quantities of relevance data were available, we could train a model to approximate the relevance probability in Equation~\ref{eq:objective} by learning a representation space where learner inputs and relevant exercises are considered similar and then performing direct search. However, input relevance data is unlikely to be available at the scale necessary to train such a model. Instead, we propose \methodname, visualized in Figure~\ref{fig:method}, which after a multilingual contrastive training stage, retrieves exercises in a two-step manner. First, we sample a set of $N_c$ hypothetical exercises from the exercise distribution \emph{conditioned on the learner input}. We call these sampled exercises our \emph{retrieval candidates}. In principle, we do not have access to this exact distribution, but we can efficiently approximate sampling via LLM. Second, we use the retrieval candidates to perform similarity search via $K$-nearest neighbors. \methodname is inspired by two complementary methods: the multilingual contrastive learning approach of \cite{wang2022english}, and the HyDE retrieval method of~\cite{gao2023precise}. We now discuss both the training and retrieval stages in greater detail.\\

\noindent \textbf{Stage 1: Learning a multilingual similarity space.}
While we operate in a setting where no explicit learner relevance data is provided, the multilingual nature of our exercises implies that a certain structure should exist in our representation space. Namely, the sentence $\exfl{i} $ in L1 should be similar to its translation $\exll{i}$ in L2. To ensure this structure is reflected in our representation space, we take inspiration from~\cite{wang2022english} and utilize multilingual contrastive learning, an unsupervised approach that aims to learn a representation where similar items (called \emph{positive} pairs) are closer together and dissimilar items (called \emph{negative} pairs) are far apart. For exercise $\ex{i}$, the contrastive loss $\calL_i$ with a mini-batch of $N_B$ sentence pairs is
\begin{align}
    \calL_i = -\log \frac{\exp\left(\text{sim}\left(\encoder(\exfl{i}), \encoder(\exll{i})\right) / \tau\right)}
    {\sum_{j=1}^{N_B}\exp\left(\text{sim}\left(\encoder(\exfl{i}), \encoder(\exll{j})\right) / \tau\right)},
\end{align}
where $\tau$ is the user-set temperature parameter and $\text{sim}\left(\cdot, \cdot\right)$ is the cosine similarity:
\begin{align}
     \text{sim}\left(\encoder(t_1), \encoder(t_2)\right) = \frac{\langle \encoder(t_1)\encoder(t_2)\rangle}{\norm{\encoder(t_1)}_2\norm{\encoder(t_2)}_2}.
\end{align}
In this work, rather than train a sentence encoder from scratch, we follow the commonly accepted practice of initializing our encoder with existing \bert-based checkpoints and contrastively finetuning these checkpoints on exercise data.\\

\noindent \textbf{Stage 2: Sampling retrieval candidates and exercise retrieval.}
A core component of \methodname is sampling from the exercise distribution conditioned on the learner input. While we cannot sample directly from this distribution, we can approximate sampling with a LLM. In particular, we prompt the LLM with a fixed a description of the exercise distribution and instruct the LLM to synthesize \emph{hypothetical} exercises based on this description and based on a learner's input. Crucially, we can synthesize exercises \emph{without requiring any labeled examples}, i.e., we do not embed examples of inputs and relevant exercises in the prompt. To retrieve exercises, the LLM synthesizes $K_h$ hypothetical exercises, which we denote $\exh{1}, \ldots, \exh{K_h}$. We then encode these hypothetical exercises via $\encoder$ to obtain $K_h$ vectors $\encoder(\exh{1}), \ldots, \encoder(\exh{K_h})$. To retrieve exercises, we retrieve the $K$ exercises that have the highest similarity score compared to the average of the $K_h$ vectors: $\frac{1}{K_h}\sum_{i=1}^{K_h} \encoder(\exh{i})$. We use GPT-4~\cite{openai2023gpt4} in this work, but in practice, any LLM of sufficient capacity can be used.

\subsection{Bridging the referential similarity gap with \methodname.}\label{sec:prob_setup:challenges}
In an effort to better understand the task of retrieving exercises from learner inputs, we crowdsourced a small dataset of learner inputs from users of \duolingo, a popular language learning app. We then contrastively finetune \mbert with roughly 40,000 real exercises from the app, spanning 4 different language courses. To get a sense of how contrastively learned similarity spaces reflect learner inputs and exercise text, we visualize our collected data, along with a subsample of the exercises, via TSNE in Figure~\ref{fig:emb-vis-final}. This visualization reveals \textbf{a fundamental referential similarity gap between learner inputs and exercise text}: How learners describe what they want to learn occupies a distinct part of the representation space, characterized by explicit use of words or phrases about language (e.g., ``verbs'', ``past tense'', ``adjectives''). We refer to this region as the ``language about language'' region. As a result, direct similarity search yields exercises that similarly contain words explicitly about language. As shown in Figure~\ref{fig:emb-vis-final}, the input ``past tense verbs'' is most similar to exercises about language (e.g., ``I explained the new \underline{words} to him''). Figure~\ref{fig:emb-vis-final} also highlights how synthesizing retrieval candidates helps bridge this referential similarity gap by ``translating'' the learner’s input (which is typically expressed in ``language about language'') to a hypothetical in-distribution exercise whose content satisfies the learner input. We provide concrete examples of learner inputs and synthesized retrieval candidates in Table~\ref{tab:syn_ex}.

\begin{table*}[t]
    \centering
    \caption{Examples of collected learner inputs and retrieval candidates synthesized based on the learner input via GPT-4. For a variety of input types, GPT-4 is able to bridge the referential similarity gap by synthesizing text that closely resembles real exercise text while incorporating the concept that the learner wants to learn.}
    \begin{tabular}{l l l}
        \toprule
        Input & Synthesized retrieval candidates & \\
        \midrule
        Past tense
        & \begin{tabular}{@{}l@{}} 
        They went to the concert last night. \\
        Did you finish your project on time? \\
        We didn't have any coffee this morning. 
        \end{tabular}
        & \begin{tabular}{@{}l@{}} 
        She cooked a delicious meal for us.\\
        He had never seen such a beautiful sunset.\\
        Were they able to solve the problem?
        \end{tabular}\\
        \midrule
        Future tense
        &\begin{tabular}{@{}l@{}}
        She will be moving to France next year. \\
        I won't attend the party tonight. \\
        When will you finish the project?
        \end{tabular}
        & \begin{tabular}{@{}l@{}}
        They'll be studying for the exam tomorrow. \\
        In five years, I'll have my own business. \\
        We're going to plant a garden this summer.
        \end{tabular} \\
        \midrule
        \begin{tabular}{@{}l@{}}
        Present \\ progressive \\ verbs
        \end{tabular}
        &\begin{tabular}{@{}l@{}}
        Are you studying for the test? \\
        She's preparing dinner for tonight. \\
        They're practicing their dance routine. \\
        \end{tabular}
        & \begin{tabular}{@{}l@{}}
        He's not listening to the lecture. \\
        I'm writing a letter to my friend. \\
        The cat is chasing its tail.
        \end{tabular} \\
        \midrule
        \begin{tabular}{@{}l@{}}
        Idiomatic \\ syntax
        \end{tabular}
        &\begin{tabular}{@{}l@{}}
        It's raining cats and dogs! \\
        Don't put all your eggs in one basket.\\
        He's feeling under the weather.
        \end{tabular}
        & \begin{tabular}{@{}l@{}}
        She has a heart of gold. \\
        I'm on cloud nine. \\
        Keep your chin up!
        \end{tabular} \\
        \midrule
        \begin{tabular}{@{}l@{}}
        How to order \\ food at a \\ restaurant
        \end{tabular}
        &\begin{tabular}{@{}l@{}}
        Could I see the menu, please? \\
        I'd like to order the grilled salmon. \\
        Does this dish contain any nuts? 
        \end{tabular}
        & \begin{tabular}{@{}l@{}}
        May I have a glass of water? \\
        Can I substitute fries for a salad? \\
        Are there any vegetarian options? \\
        \end{tabular} \\
        \bottomrule
    \end{tabular}
    \label{tab:syn_ex} 
\end{table*}

\section{Datasets and experimental results}\label{sec:exp}
In this section, we first give an overview of two novel datasets specifically for the task of learner directed language learning. We then compare \methodname against a variety of baselines on these datasets.

\subsection{Datasets}\label{sec:exp:datasets}
\noindent \textbf{Duolingo Relevance (DuoRD) Dataset.}
To evaluate our method, we collected a small scale dataset of 61 learner inputs from learners of \duolingo, a popular language learning app. For each input, we asked the learner to rate 15 exercises as relevant or irrelevant to their input, resulting in 915 total exercises rated. Exercises were sourced a pool of approximately 40,000 sentence pairs across four distinct courses. To ensure that the dataset was not skewed too heavily towards relevant or irrelevant responses, we utilize a sampling approach. Using \methodname, we retrieved the top 555 exercises in terms of similarity. To form the set of 15 exercises for the learner to rate, we select the top 5 scoring exercises deterministically (Tier 1). From the next 50 highest scoring exercises, we randomly select 5 exercises uniformly at random without replacement (Tier 2). We repeat this sampling again, randomly drawing 5 exercises from the remaining 500 exercises (Tier 3). We observe that 64\% of exercises from Tier 1 were rated as relevant, 50\% from Tier 2, and 34\% from Tier 3, resulting in 49\% of all exercises rated as relevant. \\

\noindent \textbf{Tatoeba Tags dataset.}
To test our method on a larger scale, we construct a retrieval dataset from Tatoeba, a public database of sentences and their translations. We begin by noting that when sentences are uploaded to Tatoeba, they are often tagged by grammatical concepts, language specific concepts, or topics. For example, the sentence ``The brown bear is an omnivore'' is tagged with ``animals'' and the sentence ``That way I kill two birds with one stone'' is tagged with ``idiomatic expression''. We treat each of these tags as a learner input, and deem an exercise relevant if it has been tagged accordingly. While per sentence tags are not necessarily exhaustive, they provide useful signal for evaluating retrieval approaches with typical retrieval metrics as well as binary classification metrics, as we discuss in the Section~\ref{sec:exp:metrics}. We form 3 benchmarks for evaluation, collectively referred to as the Tatoeba Tags dataset:
\begin{itemize}
    \item English benchmark: only English sentences with 139 tags and 89,392 sentences.
    \item Spanish from English benchmark: Spanish-English sentence pairs with 114 tags in Spanish and 49,258 pairs.
    \item English from Spanish benchmark: Spanish-English sentence pairs with 108 tags in Spanish and 46,837 pairs.
\end{itemize}

To form the benchmarks, we collect all tags corresponding to the benchmark, filter out tags and sentences containing profanity, merge similar tags together, and then perform benchmark specific language and content processing. We then keep only the tags with more than 20 sentences and download the corresponding sentences. The benchmark specific processing is done to better align the benchmark with how learners would interact with real-world language learning courses. Specifically, we perform both language and content processing. For language processing, we translate all tags (which appear in a variety of languages) to the L1. This is done to emulate the learning process: we use tags as a stand-in for learner inputs, which are likely to be the learner's L1. For content processing, we remove tags that do not make sense in the context of a particular learning direction. For example, a Spanish speaker learning English would not input ``voseo'' (a Spanish grammatical concept), nor would an English speaker learning Spanish input ``British English''. 

\subsection{Evaluation procedure and metrics}\label{sec:exp:metrics}
For the \duord, we treat the 915 exercises that have been rated for some learner input as the exercise set. Because each of the 915 exercises was not assigned a relevance rating for every learner inputs, we cannot use typical information retrieval metrics such as Recall or Precision. As a result, we treat evaluation on this dataset as a binary classification problem, where the goal is to predict whether an exercise is relevant or irrelevant. To evaluate methods, we use area under the receiver operating characteristic curve (AUC) and accuracy. To compute AUC, for each retrieved exercise, we compute a \textit{score} equal to the similarity measure between the retrieval candidate and all exercises. We then aggregate relevance labels and scores across all learner inputs to define the ROC curve. To compute accuracy, we compute the scores as in AUC, and set a threshold such that any exercise above the threshold is deemed relevant and vice versa. Because the similarity score ranges between -1 and 1, we set the threshold by sweeping over $[-1, 1)$ in increments of $0.1$. We then report the highest accuracy among all thresholds in the sweep.

For the Tatoeba Tags dataset, because we have a notion of relevance, as indicated by the presence of a tag, we utilize Precision@$K$, which is a common metric in information retrieval that reports the fraction of the $K$ retrieved exercises that are relevant. To compute Precision@$K$, we retrieve $K$ sentences per learner input (i.e., tag) and record the fraction of the $K$ retrieved sentences tagged with the learner input tag. Because the tagging of Tatoeba sentences is not exhaustive, the absolute values of reported Precision@$K$ may be low, but relative performance still indicates how methods would perform if tagging was comprehensive. In light of this, we again follow the evaluation approach of the \duord and report AUC. 

When performing evaluation in both datasets, we can retrieve exercises in two distinct ways. We can synthesize retrieval candidates in the L1 and perform similarity search on the L1 exercise texts. Alternatively, we can synthesize retrieval candidates in the L2 and perform similarity search on the L2 exercise texts (example translations). As a result, we report AUC, accuracy, and precision@$K$ in both the L1 and L2 setting.

\subsection{Baselines}\label{sec:exp:baselines}
For both the \duord and Tatoeba tags dataset, we evaluate \methodname against direct similarity search using \bert and \mbert~\cite{devlin2019bert}, as well as the following \bert-based models: \contriever, \mcontriever~\cite{izacard2022unsupervised}, and \simcse~\cite{gao2021simcse}. In particular, we use the \bert$_\texttt{base}$ (110 million parameters) variant of each of the above methods. These methods achieve strong unsupervised performance in a variety of retrieval and semantic text similarity tasks. \bert and \mbert were trained in a self-supervised manner by using masked language modeling and next sentence prediction objectives, with the only difference being the training data (only English for \bert and a multilingual corpus for \mbert). 

\contriever and \mcontriever propose two new approaches in contrastively tuning \bert: (1) utilizing an inverse-cloze task and independent cropping as means of forming positive pairs and (2) utilizing a Momentum encoder as described in~\cite{izacard2022unsupervised}  to ensure better representation of negative items. \contriever is initialized with \bert and trained on English CCNet~\cite{wenzek2020ccnet} and Wikipedia data, whereas \mcontriever was initialized with \mbert and trained on multiple languages in CCNet. We also consider supervised variants of \contriever and \mcontriever, which are finetuned on the MS MARCO~\cite{bajaj2016ms}, a large scale retrieval dataset. \simcse uses dropout to create synthetic positive pairs for the contrastive loss by passing the same sentence through the encoder with different random dropout parameters. Starting with \bert, \simcse is trained on Wikipedia data.

\begin{table*}[!htbp]
    \centering
    \caption{Examples of exercises retrieved with direct similarity search and \methodname for the same input on the Tatoeba Tags English Benchmark. Direct similarity search is not capable of bridging the fundamental referential similarity gap between learner inputs and exercise content, as illustrated by ``Subject verb agreement'', ``Second person'', and ``Colloquial'' inputs. In settings where learners ask about specific topics, direct similarity search returns exercises that most literally match the learner input, as shown with the ``Preference'', ``Cooking'' and ``Sports'' inputs. On the other hand, \methodname retrieves exercises well aligned with the learner input.}
    \begin{tabular}{l l l}
        \toprule
        Input & Direct similarity search & \methodname \\
        \midrule
        \begin{tabular}{@{}l@{}} Subject verb \\ agreement \end{tabular} 
        & \begin{tabular}{@{}l@{}} Correct the underlined words \\ That's a transitive verb \\ It's a transitive verb \end{tabular}
        & \begin{tabular}{@{}l@{}} The dogs are in the garden \\ They grow flowers in the garden \\ The children are playing in the garden \end{tabular} \\
        \midrule
        Second person 
        & \begin{tabular}{@{}l@{}} 
        It's secondhand \\
        It is secondhand \\
        Next person please\\
        \end{tabular}
        & \begin{tabular}{@{}l@{}} 
        Are you sure you want me to help you with your homework?\\
        I'm assuming you could speed through your testimony...\\
        Will you please check to see if my order has been dealt with?
        \end{tabular} \\
        \midrule
        Colloquial 
        & \begin{tabular}{@{}l@{}} Be punctual \\ Speaking \\ Talk is cheap \end{tabular}
        & \begin{tabular}{@{}l@{}} You drive me round the bend \\ You're laying it on a bit thick \\ You're joshing me \end{tabular} \\
        \midrule
        Preference 
        & \begin{tabular}{@{}l@{}} Make your choice \\ Compromise \\ Make a choice \end{tabular}
        & \begin{tabular}{@{}l@{}} Which do you like better, Mexican food or Chinese food? \\ Which sweet do you prefer? \\ Which do you better, pizza or tacos? \end{tabular} \\
        \midrule
        Cooking
        & \begin{tabular}{@{}l@{}} My hobby is cooking \\ Eat and drink \\ Do the laundry \end{tabular}
        & \begin{tabular}{@{}l@{}} Pour melted butter over the popcorn \\ Add the chives and season the salad \\ Will you warm up the soup? \end{tabular} \\
        \midrule
        Sports 
        & \begin{tabular}{@{}l@{}} I like playing sports \\ I love sports \\ I like sports \end{tabular}
        & \begin{tabular}{@{}l@{}} One must practice every day in order to become a worldclass athlete \\ Which do you like better skating or skiing? \\ Which do you like better cycling or jogging? \end{tabular} \\
        \bottomrule
        \\
        \\
    \end{tabular}
    \label{tab:examples} 
\end{table*}

    \begin{figure*}
    \includegraphics[width=0.9\linewidth]{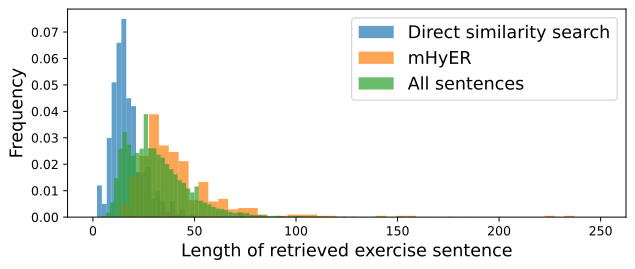}
    \captionof{figure}{Length of the top 3 retrieved exercise sentences, measured in number of characters, for direct similarity search and \methodname. Exercises retrieved via direct similarity search are inherently \emph{biased} in length, with a majority of exercises being relatively short. Using \methodname results in exercises of more varied length. This variation in length aligns well with the global distribution of exercises, showing that \methodname effectively translates learner inputs to the in-distribution exercises.} 
    \label{fig:counts}
    \end{figure*}

\subsection{Direct similarity search vs. \methodname: A qualitative case study}
Before we present our full experimental results, we first present examples of inputs and retrieved sentences on the English benchmark of the Tatoeba Tags dataset. To qualitatively gauge the difference between direct similarity search and \methodname, we provide examples of retrieved exercises for a small number of inputs in Table~\ref{tab:examples}. We present the top three retrieved exercises measured in terms of similarity score for both direct similarity search and \methodname, using \mbert finetuned on Tatoeba data as our similarity space. The input ``Subject verb agreement'' highlights the ``language about language'' phenomena: Instead of retrieving exercises containing correct subject verb agreement, direct similarity retrieves exercises in the ``language about language'' part of the similarity space. These exercises contain words such as ``words'' and ``verb''. On the other hand, \methodname is capable of bridging the gap between input and exercises, retrieving exercises that focus on ensuring sentences with plural objects have the right verb form. The ``Preference'' input illustrates an example of a nebulous input, as the learner wants exercises that have to do with expressing preferences. However, direct similarity search returns exercises explicitly about ``choice'', whereas \methodname retrieves exercises that have learners practice expressing preference in more natural settings. The last two inputs, ``Cooking'' and ``Sports'', illustrate instances where direct similarity search yields exercises that too literally match the input.

Aside from retrieval quality, we observe that retrieved results from direct similarity search also suffer from \emph{sentence length bias}. In contrastively learned similarity spaces, it has been empirically observed that the length of a sentence is implicitly encoded in the representation of a sentence, meaning sentences of a similar length are more likely to be considered similar~\cite{wu2022esimcse}. The retrieved exercises from direct similarity search shown in Table~\ref{tab:examples} clearly exhibit this bias whereas those retrieved via \methodname exhibit a higher variation in length. We confirm that this phenomena holds for all inputs in the Tatoeba Tags English benchmark by retrieving the top 3 exercises across all 139 tags with both direct similarity search and \methodname. For each exercise, we record its length (measured in number of characters). As shown in Figure~\ref{fig:counts}, exercises retrieved with \methodname are longer on average, aligning remarkably well with the global sentence length distribution. On the other hand, direct similarity search yields sentences that are notably shorter on average. This empirical observation highlights that generating in-distribution retrieval candidates allows us to retrieve sentences of varied length that track well with our set of exercises.

\subsection{Experimental results}

\begin{table*}[t]
    \centering
    \caption{Evaluation results on the \duord. $\methodname_{[\texttt{model}]}$ indicates that contrastive finetuning was employed with [$\texttt{model}$] as the initial checkpoint. $+\duord$ denotes that the \duord was used for contrastive finetuning. In all cases, \methodname outperforms relevant baselines dramatically.}
    \label{tab:duord} 
    \begin{tabular}{l l c c c c}
        \toprule
        & 
        & \begin{tabular}{@{}c@{}} AUC \\ 
        \texttt{L1}\end{tabular} 
        & \begin{tabular}{@{}c@{}}AUC \\ 
        \texttt{L2}\end{tabular} 
        & \begin{tabular}{@{}c@{}}Accuracy @ threshold\\ 
        \texttt{L1} \end{tabular} 
        & \begin{tabular}{@{}c@{}}Accuracy @ threshold\\ \texttt{L2}\end{tabular}
        \\
        \midrule
        \multirow{10}*{\begin{tabular}{@{}l@{}l@{}}
           \aptLtoX[graphic=no,type=html]{Unsupervised\,}{\rotatebox[origin=c]{90}{Unsupervised}} 
           & \aptLtoX[graphic=no,type=html]{pretraining}{\rotatebox[origin=c]{90}{pretraining}}
       \end{tabular}}
        & \bert & 0.458 & 0.421 & 0.491 @ -1.0 & 0.507 @ 0.9\\
        & \mbert  & 0.485 & 0.471 & 0.491 @ -1.0 & 0.491 @ -1.0 \\
        & \contriever & 0.565 & 0.586 & 0.557 @ 0.4 & 0.540 @ 0.3 \\
        & \mcontriever & 0.497 & 0.499 & 0.548 @ 0.6 & 0.523 @ 0.5\\
        & \simcse & 0.579 & 0.536 & 0.564 @ 0.5 & 0.523 @ 0.4\\
        & \begin{tabular}{@{}l@{}}
             $\methodname_{\mbert}$\\
             \quad +\texttt{\duord}
        \end{tabular} & 0.679 & 0.669 & 0.631 @ 0.5 & 0.605 @ 0.5\\
        & \begin{tabular}{@{}l@{}}
             $\methodname_{\mcontriever}$\\
             \quad +\texttt{\duord}
        \end{tabular} & \underline{\textbf{0.680}} & \underline{\textbf{0.678}} & \underline{\textbf{0.635 @ 0.3}} & \underline{\textbf{0.624 @ 0.3}}\\
        %
        \midrule
        \multirow{6}*{\begin{tabular}{@{}l@{}l@{}}
           \aptLtoX[graphic=no,type=html]{Supervised\,}{\rotatebox[origin=c]{90}{Supervised}} 
           & \aptLtoX[graphic=no,type=html]{pretraining}{\rotatebox[origin=c]{90}{pretraining}}
       \end{tabular}}
        & \contriever-\texttt{sup} & 0.609 & 0.605 & 0.579 @ 0.3 & 0.581 @ 0.2\\
        & \mcontriever-\texttt{sup} &  0.558 & 0.543 & 0.541 @ 0.3 & 0.529 @ 0.3\\
        & \begin{tabular}{@{}l@{}}
             $\methodname_{\contriever-\texttt{sup}}$\\
             \quad +\texttt{\duord}
        \end{tabular} & 0.674 & 0.578 & 0.632 @ 0.4 & 0.562 @ 0.4\\
        & \begin{tabular}{@{}l@{}}
             $\methodname_{\mcontriever-\texttt{sup}}$\\
             \quad +\texttt{\duord}
        \end{tabular} & \underline{\textbf{0.684}} & \underline{\textbf{0.678}} & \underline{\textbf{0.637 @ 0.3}} & \underline{\textbf{0.612 @ 0.4}}\\
        \bottomrule
    \end{tabular}
\vspace*{10pt}
\end{table*}

\begin{table*}[t]
    \centering
    \caption{Evaluation results on the Tatoeba Tags dataset. $\methodname_{[\texttt{model}]}$ indicates that contrastive finetuning was employed with [$\texttt{model}$] as the initial checkpoint. $+[\texttt{dataset name}]$ denotes that [\texttt{dataset name}] data was used for contrastive finetuning. In all cases, \methodname outperforms relevant baselines dramatically, with large gains coming from finetuning on \emph{out-of-distribution} data.}
    \label{tab:tatoeba} 
    \begin{tabular}{l l c c c c c c c c c c}
        \toprule
        \multicolumn{2}{c}{}
        & \multicolumn{2}{c}{English}
        & \multicolumn{4}{c}{English (L2) from Spanish (L1)}
        & \multicolumn{4}{c}{Spanish (L2) from English (L1)} \\ 
        & 
        & AUC 
        & P@15 
        & \begin{tabular}{@{}c@{}}AUC \\ \texttt{L1} \end{tabular} 
        & \begin{tabular}{@{}c@{}}AUC \\ \texttt{L2} \end{tabular}
        & \begin{tabular}{@{}c@{}}P@15 \\ \texttt{L1} \end{tabular} 
        & \begin{tabular}{@{}c@{}}P@15 \\ \texttt{L2} \end{tabular}
        & \begin{tabular}{@{}c@{}}AUC \\ \texttt{L1} \end{tabular} 
        & \begin{tabular}{@{}c@{}}AUC \\ \texttt{L2} \end{tabular}
        & \begin{tabular}{@{}c@{}}P@15 \\ \texttt{L1} \end{tabular} 
        & \begin{tabular}{@{}c@{}}P@15 \\ \texttt{L2} \end{tabular}
        \\
        \midrule
        \multirow{15}*{\begin{tabular}{@{}l@{}l@{}}
           \aptLtoX[graphic=no,type=html]{Unsupervised\,}{\rotatebox[origin=c]{90}{Unsupervised}} 
           & \aptLtoX[graphic=no,type=html]{pretraining}{\rotatebox[origin=c]{90}{pretraining}}
       \end{tabular}}
        & \bert & 0.495 & 0.032 & 0.481 & 0.428 & 0.019 & 0.020 & 0.492 & 0.505 & 0.044 & 0.020 \\
        & \mbert & 0.468 & 0.037 & 0.446 & 0.487 & 0.038 & 0.040 & 0.469 & 0.442 & 0.039 & 0.019 \\
        & \contriever & 0.536 & 0.161 & 0.542 & 0.523 & 0.112 & 0.073 & 0.529 & 0.549 & 0.165 & 0.087 \\
        & \mcontriever & 0.571 & 0.064 & 0.438 & 0.503 & 0.051 & 0.063 & 0.559 & 0.564 & 0.061 & 0.027 \\
        &\simcse & 0.646 & 0.115 & 0.535 & 0.559 & 0.069 & 0.054 & 0.635 & 0.610 & 0.127 & 0.068 \\
        & \begin{tabular}{@{}l@{}}
             $\methodname_{\mbert}$\\
             \quad +\texttt{en-from-es}
        \end{tabular} & 0.722 & 0.225 & 0.686 & 0.701 & 0.227 & 0.208 & 0.710 & 0.696 & 0.243 & 0.242 \\
        & \begin{tabular}{@{}l@{}}
             $\methodname_{\mbert}$\\
             \quad +\texttt{es-from-en}
        \end{tabular} & 0.717 & 0.223 & 0.697 & 0.693 & 0.219 & 0.211 & 0.702 & 0.706 & 0.237 & 0.244 \\
        & \begin{tabular}{@{}l@{}}
             $\methodname_{\mbert}$\\
             \quad +\texttt{\duord}
        \end{tabular} & 0.752 & 0.211 & 0.734 & 0.738 & 0.215 & 0.206 & 0.739 & \underline{\textbf{0.757}} & 0.225 & 0.242 \\
        & \begin{tabular}{@{}l@{}}
             $\methodname_{\contriever}$\\
             \quad +\texttt{\duord}
        \end{tabular} & \underline{\textbf{0.768}} & 0.239 & 0.644 & \underline{\textbf{0.780}} & 0.106 & 0.232 & \underline{\textbf{0.749}} & 0.659 & 0.265 & 0.099 \\
        & \begin{tabular}{@{}l@{}}
             $\methodname_{\mcontriever}$\\
             \quad +\texttt{\duord}
        \end{tabular} & 0.729 & \underline{\textbf{0.258}} & \underline{\textbf{0.748}} & 0.723 & \underline{\textbf{0.267}} & \underline{\textbf{0.264}} & 0.713 & 0.744 & \underline{\textbf{0.271}} & \underline{\textbf{0.294}} \\
        \midrule
        \multirow{6}*{\begin{tabular}{@{}l@{}l@{}}
           \aptLtoX[graphic=no,type=html]{Supervised\,}{\rotatebox[origin=c]{90}{Supervised}} 
           & \aptLtoX[graphic=no,type=html]{pretraining}{\rotatebox[origin=c]{90}{pretraining}}
       \end{tabular}}
        & \contriever-\texttt{sup} & 0.541 & 0.164 & 0.491 & 0.492 & 0.120 & 0.086 & 0.530 & 0.492 & 0.180 & 0.105 \\
        & \mcontriever-\texttt{sup} & 0.575 & 0.104 & 0.548 & 0.510 & 0.126 & 0.108 & 0.560 & 0.581 & 0.112 & 0.101 \\ 
        & \begin{tabular}{@{}l@{}}
         $\methodname_{\contriever-\texttt{sup}}$\\
         \quad +\texttt{\duord}
        \end{tabular} & \underline{\textbf{0.775}} & 0.246 & 0.668 & \underline{\textbf{0.797}} & 0.102 & 0.240 & \underline{\textbf{0.760}} & 0.692 & \underline{\textbf{0.268}} & 0.108 \\ 
        & \begin{tabular}{@{}l@{}}
             $\methodname_{\mcontriever-\texttt{sup}}$\\
             \quad +\texttt{\duord}
        \end{tabular} & 0.738 & \underline{\textbf{0.255}} & \underline{\textbf{0.761}} & 0.734 & \underline{\textbf{0.260}} & \underline{\textbf{0.264}} & 0.722 & \underline{\textbf{0.752}} & 0.255 & \underline{\textbf{0.280}} \\ 
        \bottomrule
    \end{tabular}
    \vspace{-0.75\baselineskip}
\end{table*}

In this section, we present our experimental results on the \duord and Tatoeba Tags dataset. For both datasets, we consider two starting points for fine-tuning the \bert embedding model: \emph{Unsupervised pretraining}, where we contrastively train a \bert checkpoint that has been pretrained in an unsupervised manner, and \emph{supervised pretraining}, where we start with a \bert checkpoint that has been pretrained on MS MARCO~\cite{bajaj2016ms}, a large scale retrieval dataset that covers different tasks, such as passage ranking and keyphrase extraction. In all settings, $\methodname_{[\texttt{model}]}$ denotes \methodname with starting with [\texttt{model}] as its initial checkpoint for contrastive training. [\texttt{model}]-\texttt{sup} indicates [\texttt{model}] was trained in a supervised manner. \textbf{We emphasize that at no point in training \methodname is labeled training data for exercise retrieval used}; \texttt{sup} only indicates MS MARCO was used to train the initial \bert checkpoint. For all experiments, we take the $\texttt{[CLS]}$ representation as the sentence representation, except when working with \contriever and \mcontriever, where we use their custom mean pooling\footnote{See \url{https://huggingface.co/facebook/contriever} for further details.}. For all experiments with \methodname, we adopt the training setup of~\cite{wang2022english}, which is adapted from~\cite{gao2021simcse}, including all default hyperparameters. For retrieval, we synthesize $K_h = 10$ hypothetical retrieval candidates from GPT-4 and perform nearest neighbors search with the averaged embedding of all $K_h$ candidates.\\

\noindent \textbf{\duord.} The evaluation results of baselines and \methodname on the \duord are presented in Table~\ref{tab:duord}. For both unsupervised and supervised settings, we contrastively finetune \bert checkpoints on the full 40K exercises in the \duord. In the unsupervised pretraining setting, we start our contrastive finetuning with two multilingual checkpoints: \mbert and \mcontriever. In this setting, \methodname outperforms all relevant baselines in both AUC and accuracy, with \methodname${_\mcontriever}$ achieving the best performance among all methods. \textbf{\methodname${_\mcontriever}$ results in 36.8\% and 40.2\% AUC gains over \mcontriever and \mbert, respectively.} It is notable that direct similarity search generally fails to perform well, highlighting that the gap between learner inputs and relevant exercises: \bert, \mbert, and \mcontriever baselines fail to even achieve an AUC of 0.5 corresponding to random guessing, reinforcing the fact that \textbf{direct similarity search cannot overcome the fundamental mismatch between how learners describe what they want to learn and exercise content.}
In the supervised pretraining setting, we start our finetuning from the \contriever-\texttt{sup} and \mcontriever-\texttt{sup} checkpoints, which were finetuned on labeled MS MARCO data. \methodname once again outperforms all baselines, with \methodname$_{\mcontriever-\texttt{sup}}$ as the best performing method. Here, supervised pretraining modestly improves the performance of direct similarity search, suggesting that supervised pretraining can lessen the referential similarity gap in a limited manner. The improvement due to supervised pretraining is less pronounced when utilizing \methodname, with even one instance of decreased accuracy. This suggests synthesized retrieval candidates bridge the gap to the point where further improvement is difficult. \\ 

\noindent \textbf{Tatoeba Tags dataset.} The evaluation results of baselines and \methodname on the Tatoeba Tags dataset are presented in Table~\ref{tab:tatoeba}. On this dataset, we experiment with contrastive finetuning on \emph{out-of-distribution data}. This experiment was inspired by empirical observations from finetuning \mbert. In particular, we contrastively finetuned \mbert on the Spanish from English benchmark (denoted \texttt{es-from-en}) and the English from Spanish benchmark (denoted \texttt{en-from-es}), as well as the 40K \emph{out-of-distribution} sentence pairs from the \duord (which contains English-Spanish pairs). We observe that finetuning on the \duord outperforms finetuning on in-distribution data. This surprising observation leads us to finetune \contriever and \mcontriever checkpoints with the \duord in both the unsupervised and supervised settings. In the unsupervised setting, we once again observe poor performance from direct similarity search baselines and sizable increases in performance when using \methodname: \textbf{Up to 39\% increases in AUC and more than doubling the performance of precision@15 between the best \methodname method and best direct similarity.} We observe similar gains in the supervised pretraining setting.
Methods that use \contriever (pretrained only on English data) typically perform better when retrieving in English, whereas methods using \mcontriever typically perform better in multilingual settings. 

\subsection{Ablation study}
\begin{table*}[t]
    \centering
     \caption{Ablation results on the Tatoeba Tags dataset. We experiment by removing either the contrastive finetuning step or the retrieval candidate synthesis step. +\texttt{GPT} indicates that retrieval candidates were used with no contrastive finetuning, whereas +\duord indicates that direct similarity search was used after contrastively finetuning on the \duord. In a vast majority of cases, contrastive finetuning and retrieval candidate synthesis boost performance, with retrieval candidates generally contributing more.}
    \label{tab:ablations:tatoeba}
    \begin{tabular}{l l c c c c c c c c c c}
        \toprule
        \multicolumn{2}{c}{}
        & \multicolumn{2}{c}{English}
        & \multicolumn{4}{c}{English (L2) from Spanish (L1)}
        & \multicolumn{4}{c}{Spanish (L2) from English (L1)} \\ 
        & 
        & AUC 
        & P@15 
        & \begin{tabular}{@{}c@{}}AUC \\ \texttt{L1} \end{tabular} 
        & \begin{tabular}{@{}c@{}}AUC \\ \texttt{L2} \end{tabular}
        & \begin{tabular}{@{}c@{}}P@15 \\ \texttt{L1} \end{tabular} 
        & \begin{tabular}{@{}c@{}}P@15 \\ \texttt{L2} \end{tabular}
        & \begin{tabular}{@{}c@{}}AUC \\ \texttt{L1} \end{tabular} 
        & \begin{tabular}{@{}c@{}}AUC \\ \texttt{L2} \end{tabular}
        & \begin{tabular}{@{}c@{}}P@15 \\ \texttt{L1} \end{tabular} 
        & \begin{tabular}{@{}c@{}}P@15 \\ \texttt{L2} \end{tabular}
        \\
        \midrule
        \multirow{7}*{\begin{tabular}{@{}l@{}l@{}}
           \aptLtoX[graphic=no,type=html]{Unsupervised\,}{\rotatebox[origin=c]{90}{Unsupervised}} 
           & \aptLtoX[graphic=no,type=html]{pretraining}{\rotatebox[origin=c]{90}{pretraining}}
       \end{tabular}}
        & \mcontriever & 0.571 & 0.064 & 0.438 & 0.503 & 0.051 & 0.063 & 0.559 & 0.564 & 0.061 & 0.027 \\
        & \begin{tabular}{@{}l@{}}
             \mcontriever\\
             \quad +\texttt{GPT}
        \end{tabular} & 0.676 & 0.237 & 0.613 & 0.663 & 0.213 & 0.213 & 0.643 & 0.602 & 0.245 & 0.217 \\
        & \begin{tabular}{@{}l@{}}
             \mcontriever\\
             \quad +\texttt{\duord}
        \end{tabular} & 0.665 & 0.096 & 0.670 & 0.665 & 0.119 & 0.106 & 0.656 & 0.657 & 0.090 & 0.077 \\
        & \begin{tabular}{@{}l@{}}
             $\methodname_{\mcontriever}$\\
             \quad +\texttt{\duord}
        \end{tabular} & \underline{\textbf{0.729}} & \underline{\textbf{0.258}} & \underline{\textbf{0.748}} & \underline{\textbf{0.723}} & \underline{\textbf{0.267}} & \underline{\textbf{0.264}} & \underline{\textbf{0.713}} & \underline{\textbf{0.744}} & \underline{\textbf{0.271}} & \underline{\textbf{0.294}} \\
        \midrule
        \multirow{7}*{\begin{tabular}{@{}l@{}l@{}}
           \aptLtoX[graphic=no,type=html]{Supervised\,}{\rotatebox[origin=c]{90}{Supervised}} 
           & \aptLtoX[graphic=no,type=html]{pretraining}{\rotatebox[origin=c]{90}{pretraining}}
       \end{tabular}}
        & \mcontriever-\texttt{sup} & 0.575 & 0.104 & 0.548 & 0.510 & 0.126 & 0.108 & 0.560 & 0.581 & 0.112 & 0.101 \\
        & \begin{tabular}{@{}l@{}}
             \mcontriever-\texttt{sup}\\
             \quad +\texttt{GPT}
        \end{tabular} & 0.731 & 0.250 & 0.642 & 0.724 & 0.238 & 0.243 & 0.706 & 0.636 & \underline{\textbf{0.263}} & 0.258 \\
        & \begin{tabular}{@{}l@{}}
             \mcontriever-\texttt{sup}\\
             \quad +\texttt{\duord}
        \end{tabular} & 0.672 & 0.106 & 0.678 & 0.677 & 0.128 & 0.120 & 0.662 & 0.661 & 0.113 & 0.091 \\
        & \begin{tabular}{@{}l@{}}
             $\methodname_{\mcontriever-\texttt{sup}}$\\
             \quad +\texttt{\duord}
        \end{tabular} & \underline{\textbf{0.738}} & \underline{\textbf{0.255}} & \underline{\textbf{0.761}} & \underline{\textbf{0.734}} & \underline{\textbf{0.260}} & \underline{\textbf{0.264}} & \underline{\textbf{0.722}} & \underline{\textbf{0.752}} & 0.255 & \underline{\textbf{0.280}} \\
        \bottomrule
    \end{tabular}
\end{table*}

The two key steps in \methodname are multilingual contrastive pretraining and synthesizing retrieval candidates. To characterize the relative contributions of each step, we create variants of \methodname performing direct similarity search after contrastive pretraining or retrieving with GPT-synthesized retrieval candidates with a non-finetuned encoder (i.e., HyDE~\cite{gao2023precise}). As shown in Table~\ref{tab:ablations:tatoeba}, the combination of both stages yields the best performance in the vast majority of cases. Utilizing only synthesized retrieval candidates results in the larger increases in precision compared to contrastive finetuning, while the opposite is true for AUC. This suggests that the two steps drive performance increases in complementary ways: Contrastive finetuning changes the similarity space such that relevant exercises are closer to learner inputs at a \emph{global} level, resulting in increases in AUC (which measures a global ranking of predicitions). However, direct similarity search still cannot overcome referential similarity gaps, and hence, increases in precision@$15$ are low relatively. Meanwhile, synthesizing retrieval candidates directly improves retrieval quality, resulting in higher retrieval quality, but does not change representations, resulting in relatively lower increases in AUC.

\section{Discussion}\label{sec:discussion}
In this paper, we introduce the problem of exercise retrieval for learner directed language learning and highlight an important challenge in this setting: how learners express what they want to learn and exercise content are fundamentally semantically different. The effects of this referential similarity gap are especially pronounced when attempting to retrieve exercises via direct similarity search: even models supervised on MS MARCO, a large scale retrieval dataset, struggle to bridge this referential similarity gap. As a result, we propose \methodname, a zero-shot retrieval approach that leverages the generative capabilities of pretrained LLMs to synthesize relevant in-distribution sentences which are then used to retrieve exercises. We form two novel benchmark datasets by collecting human responses and processing publicly available data. \methodname outperforms several strong baselines, including ones trained in a supervised fashion. \\

\noindent \textbf{Future work.}
\methodname lays the methodological foundation for self-directed online language learning. Many interesting directions of future work exist, ranging from investigating different learning areas to methodological extensions that accommodate labeled relevance information. We discuss several of these directions below. 

\methodname provides concrete methodology that can enable future investigations into the effects of self-directed learning on long-term learning outcomes and curriculum design at scale. Self-directed learning also play a role in improving other components of personalized education systems. For example, a learner input into a self-directed learning system can be viewed as an indicator of a self-perceived weakness, which would provide a powerful form of supervision for estimating user skill levels. Studying how inputs and outputs of complementary parts of a unified personalized education system is an important direction of future work.

Another interesting avenue of future work is investigating if analogous ``language about language'' phenomena appear in settings other than language learning. We hypothesize that such phenomena exist in one form or another across all learning settings. For example, how learners describe what they want to review in math (e.g., ``right angles'') exhibits a similar fundamental mismatch with exercise text (e.g., ``Compute the length of the hypotenuse of this triangle''). If such gaps exist, methods capable of bridging the referential similarity gap, like \methodname, will be required across different learning settings. Characterizing the degree to which such gaps appear, as well as how such gaps differ,  in different learning settings remains important and open work. 

From a system design and learner experience perspective, developing machine learning methods to retrieve relevant exercises based on learner inputs is a foundational piece of any self-directed language learning system. However, serving a set of exercises that maximizes relevance may not lead to the best learner experience. Instead, the objective of exercise retrieval can be made more flexible: Instead of retrieving $K$ exercises that maximize relevance, we retrieve all exercises with scores that exceed some pre-determined threshold. Then, this set of relevant exercises can be re-ranked based on additional criteria, such as difficulty level (with information from Knowledge Tracing-based parts of the system) or diversity (in terms of difficulty or length). Regardless of precise objective (top $K$ vs. all relevant exercises), the referential similarity gap persists, making \methodname especially suitable for this initial retrieval step.

Methodologically, \methodname was designed explicitly with the goal of zero-shot retrieval. However, opportunities to collect learner relevance feedback grow as self-directed learning systems get implemented. Such feedback can then be used to train retrieval methods. Investigating how to effectively use limited learner feedback to help retrieval methods bridge the referential similarity gap remains an open question. Additionally, extensions of \methodname to learning settings with multi-modal exercises is direction of future work. Using newly developed multi-modal models to measure similarity in different domains, such as images or audio, can unlock a richer learning experience for learners.\\

\noindent \textbf{Acknowledgements.} We thank Ali Malik, Stephen Mayhew, and Mark Davenport for their helpful comments, feedback, and support. AX was partially supported by National
Science Foundation grants 
IIS-2212182 and DMS-2134037.

\balance
\bibliographystyle{ACM-Reference-Format}
\bibliography{bib}

\end{document}